\documentstyle[preprint,times,aps,prc,epsf,floats,tighten]{revtex}
\begin{document}
\preprint{SSF94-10-01 \hspace{0.5 cm}
To be published in Physics Letters B}
\title{Short-range correlations
in (e,e$'$p) and (e,e$'$pp) reactions on complex nuclei}
\author{Jan Ryckebusch \thanks{Postdoctoral Research Fellow NFWO},
Marc Vanderhaeghen \thanks{Research Fellow NFWO}, Kris Heyde and
Michel Waroquier \thanks{Research Director NFWO}}
\address{Laboratory for Theoretical Physics \protect\\Proeftuinstraat 86
\protect\\ B-9000 Gent, Belgium}
\date{\today}
\maketitle
\begin{abstract}
The influence of short-range correlations (SRC) on the
triple-coincidence (e,e$'$pp) reactions is studied.  The
non-relativistic model uses a mean-field potential to account for the
distortions that the escaping particles undergo.  Apart from the SRC,
that are implemented through a Jastrow ansatz with a realistic
correlation function, we incorporate the contribution from pion
exchange and intermediate $\Delta _{33}$ currents.  The (e,e$'$pp)
cross sections are predicted to exhibit a sizeable sensitivity to the
SRC.  The contribution from the two-nucleon breakup channel to the
semi-exclusive $^{12}$C(e,e$'$p) cross section is calculated in the
kinematics of a recent NIKHEF-K experiment.  In the semi-exclusive
channel, a selective sensitivity in terms of the missing energy and
momentum to the SRC is found.
\end{abstract}
\newpage
The (e,e$'$p) reaction has been explored extensively to study the
single-particle properties of complex nuclei.  The optimum kinematical
conditions for such investigations are those in which the residual (A-1)
nucleons are behaving as spectators in the reaction of the target
nucleus with the external
electromagnetic field.  The opposite situation in which a breakup
of the (A-1) system is observed is not only a direct indication for the
existence of correlations in nuclei but can be explored to improve
on our understanding of those correlations.  As the most important
correlations in the nucleus are believed to be of two-body nature,
two-nucleon knockout is expected to be a major source
of multi-particle breakup.

Correlations of the short-range
type have been the subject of extensive theoretical studies for many
years.  In reactions with external electromagnetic probes, short range
correlations (SRC) are
competing with correlations of longer range.  Therefore, in order to
reach a maximum sensitivity to the SRC, special care must be taken in
the choice of the kinematics.  In two-nucleon knockout
reactions induced by real photons,
correlations related to pion exchange
were shown to play a predominant role \cite{ryc94}.
In the
non-relativistic limit, however, pion exchange is not affecting the charge
operator and as such two-nucleon knockout reactions with virtual
photons, which are also probing longitudinal degrees of freedom,
are expected to open better perspectives to study SRC. On the other
hand, it should be realized that beyond the quasi-elastic peak the
longitudinal part of the
inclusive (e,e$'$) cross section was experimentally verified to be
small.  This means that in predominantly longitudinal kinematics,
(e,e$'$p...) cross sections are anticipated to be small.  Another
property that can be exploited to maximize the effect of the SRC is the
fact that pion exchange currents are of charge-exchange nature.
Consequently, the two-proton emission channel is expected to open better
perspectives in the study of SRC than proton-neutron emission.

In this paper we aim at estimating the effect of SRC for the (e,e$'$pp)
reaction on complex nuclei.  Further we investigate in how far the
semi-exclusive (e,e$'$p) channel exhibits a sensitivity to the SRC and
can be used to discriminate between the different prescriptions for the
correlation functions.  In all calculations presented here we coherently
add the contribution from the short-range effects in the  nuclear wave
functions to the contribution from the pion exchange currents.

Using standard rules, the (e,e$'$N$_a$ N$_b$) cross section can be cast in
the form :
\begin{eqnarray}
{d^8 \sigma \over dE_b d \Omega _b d \Omega _a d \epsilon ' d \Omega
_{\epsilon '}} (e,e'N_aN_b) & = &
{1 \over 4 (2\pi)^8 } k_a k_b E_a E_b f_{rec} \sigma_{M}
\\ \nonumber
& & \times \left[ v_L W_L +v_T W_T + v_S W_{TT} + v_I W_{LT}\right] \;,
\label{eepnn}
\end{eqnarray}
where  $\sigma _M$ is the Mott cross section, $f_{rec}$ a recoil factor
and the coefficients $v$ contain the electron kinematics.
The structure functions W are defined in terms of the transition
matrix elements :
\begin{equation}
m_{fi}(\lambda)= \left< \overline{\Psi} _f \mid J_{\lambda}^{[1]}
+ J_{\lambda}^{[2]} \mid \overline{\Psi}_i \right>
\hspace{1.5 cm} (\lambda=0,\pm 1)\; ,
\label{matrixel}
\end{equation}
where the virtual photon is assumed to probe both one-body
$J_{\lambda}^{[1]}$ and
two-body currents $J_{\lambda}^{[2]}$ in the target nucleus.
In the absence of correlations
beyond those implemented in the mean-field, the wave functions
$\overline{\Psi}$ reduce to Slater determinants and only two-body
currents will produce non-zero matrix elements.  Here, we consider
A-body wave functions in the context of the Correlated Basis Function
(CBF) theory \cite{benhar0,benhar} :
\begin{equation}
\overline{\Psi}= {\cal G} \mid \Psi > \;,
\end{equation}
where ${\cal G}$ projects on the Slater determinants the correlations
that are absent in an independent-particle model  (IPM) description.
The dominant terms in the  operator ${\cal G}$ are the central, or
Jastrow, component ($\prod_{i<j} f_C(r_{ij})$) and the tensor term
($\prod_{i<j} f_{t \tau} (r_{ij}) S_{ij} \vec{\tau}_i .
\vec{\tau}_j$) \cite{benhar}. Here, we retain only a state independent Jastrow
correlation term. For an initial calculation, the tensor terms can be
neglected given the smallness of the function
$f_{t \tau}$ relative to $f_C$ \cite{revmod}.  For the particular case of
two-nucleon breakup in the final state, the correlated initial and final
wave function read then :
\begin{eqnarray}
\overline{\Psi}_i & = &\prod_{i<j=1}^{A} f_C(r_{ij}) \Psi _{i}
\nonumber \\
\overline{\Psi}_f & = & \mid \vec{k}_a \frac{1}{2} m_{s_{a}},
\vec{k}_b \frac{1}{2} m_{s_{b}} >
\prod_{i<j=1}^{A-2} f_C (r_{ij}) \mid J_R M_R > \; ,
\label{corrif}
\end{eqnarray}
where $\mid J_R M_R >$ is the state in which the residual A-2 system is
created, $r_{ij}=\mid {\vec r}_i -{\vec r}_j \mid $ and
$\mid \vec{k}_a \frac{1}{2} m_{s_{a}},
\vec{k}_b \frac{1}{2} m_{s_{b}} > $ determines the momentum and spin
orientation of the ejected nucleon pair.
 Inserting the above wave functions into the matrix elements
(\ref{matrixel})  leads to an expression which cannot be calculated
exactly with the presently available computers. Various cluster
expansions, however, have been developed to calculate the matrix
elements to a high degree of accuracy \cite{pieper}. Here, we rely on a
technique developed by Weise, Huber and Danos \cite{weise}.  Essentially
it reduces the calculation of expectation values between correlated wave
functions to evaluating matrix elements between Slater determinants with
an effective operator that is constructed  from the original transition
operator and the correlation function $f_C$.  It can be readily verified
that the matrix elements of Eq.~(\ref{matrixel})
 can be rewritten as :
\begin{equation}
\left< \vec{k}_a \frac{1}{2} m_{s_{a}},
\vec{k}_b \frac{1}{2} m_{s_{b}} ; J_R M_R \mid
{\cal H}^{eff}_{\lambda} \mid \Psi _i \right> \; ,
\label{eq:trans}
\end{equation}
with
\begin{equation}
{\cal H}^{eff}_{\lambda} = \prod_{i<j=1}^{A-2}
\left( 1-g^{\dag}(r_{ij}) \right) \left( \sum_{k=1}^{A}J^{[1]}_{\lambda}(k) +
\sum_{k<l=1}^{A}J^{[2]}_{\lambda}(k,l) \right)
\prod_{m<n=1}^{A}
\left( 1-g(r_{mn}) \right) \; .
\label{exact}
\end{equation}
In the above equation we have introduced the correlation function
$g(r_{12}) \equiv 1-f_C(r_{12}$).  As outlined in Ref.~\cite{weise}, the
effective transition hamiltonian ${\cal H}^{eff}$ can be expanded into
different orders in $g$.
For the present purposes, the expansion in $g$ is expected to converge
rapidly. First, chances are relatively small that virtual photons will
interact with heavily correlated three,four, ...-nucleon clusters at
normal nuclear densities.  Further, two-body photoabsorption is
anticipated to represent the dominant absorption mechanism in
two-nucleon knockout processes, thus giving a natural constraint on the
expansion into the different orders of $g$. The convergence of the
cluster expansion in the different orders of $g(r_{12})$ has recently
been investigated for the $^4$He(e,e$'$d) case by Leidemann {\em et al.}
\cite{leideman}.  They concluded that in the (e,e$'$d)
case a first order calculation would be sufficient to incorporate the
main effects of the SRC.  The deviation between the first order and full
calculation was found to grow with increasing missing momentum $p_m$ but
remained reasonable over the whole $p_m$ range.

\begin{figure}[htb]
\centering
\epsfysize=10.cm
\epsffile{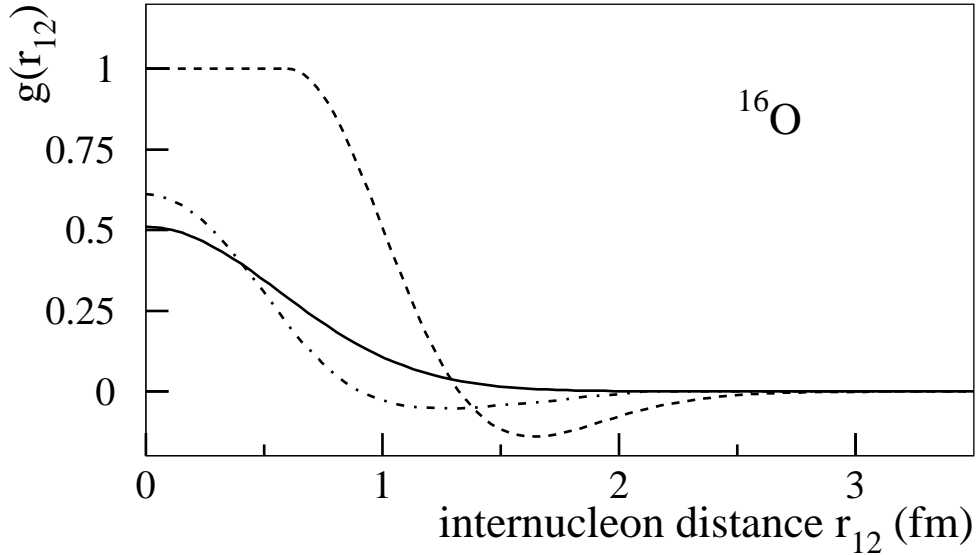}
\caption{Central correlation function for $^{16}$O as obtained in
various approximations.}
\label{srccur}
\end{figure}

Within the adopted
assumptions the effective interaction hamiltonian (\ref{exact}) reads :
\begin{equation}
{\cal H}^{eff}_{\lambda} \approx \sum_{i=1}^{A}J^{[1]}_{\lambda}(i)
-\sum_{i<j=1}^{A} \left( J^{[1]}_{\lambda}(i)
+J^{[1]}_{\lambda}(j) \right) g(r_{ij}) +
\sum_{i<j=1}^{A}J^{[2]}_{\lambda}(i,j) \left(1-g(r_{ij}) \right) \;.
\label{effh}
\end{equation}
Strictly speaking, this effective interaction hamiltonian also contains
terms in $g^{\dag}J_{\lambda}$ and $g^{\dag}J_{\lambda}g$.  All of those
terms, however, involve final-state correlations that according to
Eq.~(\ref{corrif}) refer solely to the coordinates of the A-2
spectator nucleons.  As such they do not contribute to the matrix
elements of Eq.~(\ref{eq:trans}). In deriving the expression (\ref{effh}) we
have further neglected all three- and four-body operators.  The latter
require respectively three and four active nucleons in the virtual
photon absorption process.  We expect these mechanisms to produce small
contributions in the two-nucleon breakup channel.  It remains to be
investigated whether these multi-body operators (N $\geq$ 3) give rise
to sizeable three and more nucleon knockout strength. The first term in
the effective transition hamiltonian (\ref{effh}) does not contribute to
the two-nucleon knockout channel.  The second term is a typical SRC
effect and will be referred to as the "SRC current" in the remainder of
this paper.  For the longitudinal one-body operator we consider the
charge density (neglecting the Darwin-Foldy term) $J_0 ^{[1]}=
\sum _{i=1}^{A} e_i \delta(\vec{r}-\vec{r}_i)$.  The transverse
one-body current $J_{\pm 1} ^{[1]}$ which is part of the SRC current is
constructed from the common convection and magnetization term.  The
two-body current $J_{\lambda} ^{[2]}$ is determined along the lines
explained in Ref.~\cite{ryc94}.  In short, we have considered all
diagrams with one exchanged pion including those with an intermediate
$\Delta _{33}$ excitation.  In the non-relativistic limit, the
transverse current operator that originates from this procedure has
several components including the seagull, pion-in-flight and
$\Delta$-isobar term.  The correction for the finite extension of the
interacting hadrons to
$J_{\lambda} ^{[2]}$ is handled in the standard manner by introducing a
monopole $\pi NN$ form factor with a cutoff mass $\Lambda _{\pi}=1250$
MeV/c$^2$, which is consistent with the Bonn boson exchange model for
the nucleon-nucleon interaction.  In the context of ($\gamma$,NN)
reactions, the correlation function $g(r_{ij})$ in the last term of the
interaction hamiltonian (\ref{effh}) was shown to produce an overall
reduction of the two-body current contribution $J^{[2]}_{\lambda}$ of
less than 10\% \cite{marc}.  For the calculations presented below this
effect has been neglected.

In the choice of the central correlation function $g$ we have been led
by the apparent success of the recent Variational Monte Carlo (VMC)
\cite{pieper,revmod} and Fermi Hypernetted Chain (FHNC) calculations
\cite{co} in treating complex nuclei. In the calculations of
Ref.~\cite{co} a Gaussian correlation function $g(r_{12})=\alpha
e^{-\beta r_{12}^2}$ was put forward.  For $^{16}$O the following values
for the two parameters were obtained : $\alpha$=0.51 and
$\beta$=1.52~fm$^{-2}$.  In Fig.~\ref{srccur} we compare this FHNC result
with the correlation function obtained with VMC techniques.  It is
noted that the two independent calculations produce central correlations
that bear a strong resemblance with each other.  All results presented
below are obtained with the Gaussian correlation function.  The VMC
correlation function was checked to produce very similar results.

In the calculation of the two-nucleon knockout cross sections we do not
attempt to include the full complexity of the final state interaction
and adopt a direct knockout reaction model.  This means that within our
model assumptions photoabsorption on a correlated nucleon pair does
imply that the active nucleons are ejected from the target system and
become asymptotically free particles.  The distorting effect of the
residual A-2 system on the wave functions for the escaping particles is
implemented.  This is accomplished by performing a partial wave
expansion for both of the escaping particles in terms of the
eigenfunctions of a mean-field potential  \cite{ryc93}.  In this
procedure, the initial and final state are guaranteed to remain
orthogonal, thus avoiding spurious contributions entering the matrix
elements. This is particularly of importance for the semi-exclusive
calculations, where integrations over a large fraction of phase space
are carried out.  The main effect of the distortions on the calculated
$(e,e'NN)$ cross sections is a reduction relative to the results
obtained in a plane wave approach. As such the effect of the distortions
is similar to what was observed for $(\gamma,NN)$ reactions
\cite{ryc93}.
Also in the optical potential calculations of Ref.~\cite{carfsi} the
final state interaction
was reported to bring about a moderate reduction of the two-nucleon knockout
cross sections.

\begin{figure}[htb]
\centering
\epsfysize=10.cm
\epsffile{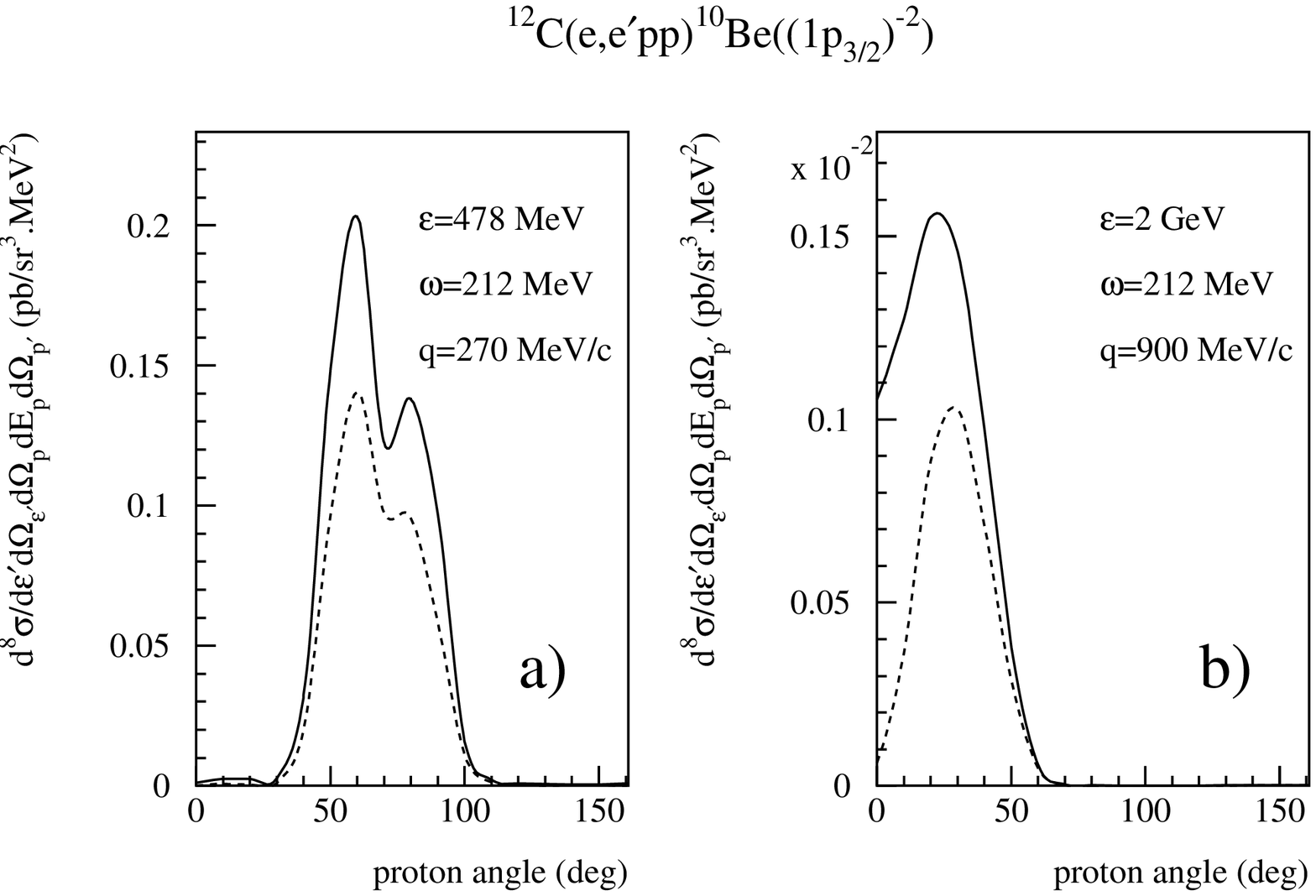}
\caption{Cross sections for the
$^{12}$C(e,e$'$pp)$^{10}$Be((1p$_{3/2})^{-2}$) reaction in coplanar and
symmetrical kinematics ($\theta _p = \theta _{p'}$).  The dashed line
shows the contribution from intermediate $\Delta _{33}$ creation.  For
the solid line the SRC term has been coherently added to the
$\Delta _{33}$ term.}
\label{eeppres}
\end{figure}

Calculated cross sections for the
$^{12}$C(e,e$'$pp)$^{10}$Be reaction leaving the residual nucleus in a
(1p$_{3/2})^{-2}$
two-hole state are shown in Fig.~\ref{eeppres}. The presented cross
sections are obtained in so-called coplanar and symmetrical kinematics
\cite{giusti1}
($\phi _p=0^o, \phi _{p'}=180^o$ and $\theta _p = \theta _{p'}$).  The
kinetic energy for one of the ejected protons was fixed such that
T$_p$=T$_{p'}$ for $\theta _p = \theta _{p'}=90^o$.  For the curves of
Fig.~\ref{eeppres} we have summed over the two possible values for the
momentum J$_R$ (0$^+$ and 2$^+$) in which the residual nucleus can be
created.   In the absence of correlations
beyond the IPM, the sole contributing two-body current to the
(e,e$'$pp) channel is of isobaric origin.  The kinematics for
Fig.~\ref{eeppres}a)
are taken from a recent NIKHEF-K $^{12}$C(e,e$'$pp) experiment \cite{leon1}.
The
curves of Fig.~\ref{eeppres}b) are obtained in more longitudinal
kinematics (the longitudinal polarization $\epsilon _L = v_L/2 v_T$ is
0.85 for kinematics b) and 0.30 for a)).  Although the effect of the SRC
is slightly bigger in more longitudinal kinematics, the predicted
increase in the cross section is less than a factor of two in the peak
of the cross section.

The (e,e$'$pp) results presented in Refs.\cite{giusti1,giusti2} predict
a much stronger sensitivity to the SRC for some particular choices of
the correlation function.  Particularly with the so-called "omy"
correlation function \cite{clark} spectacular increases in the
(e,e$'$pp) cross sections were observed.  For completeness the hard-core
 "omy" correlation function has been added to Fig.~\ref{srccur}.  The
"omy" correlation function has been derived by minimizing the
ground-state energy of $^{16}$O with a rather crude form of the
nucleon-nculeon interaction that has a state-independent hard core
radius c=0.6~fm \cite{clark}.  It is
obvious that in the "omy"  parametrization the central correlation
corrections to the wave functions are much bigger than what modern
theories predict.  A similar type of trend is observed with respect to
the effect of the SRC on the (e,e$'$pp) cross sections, confirming their
sensitivity to the choice of the correlation function.

\begin{figure}
\centering
\epsfysize=15.cm
\epsffile{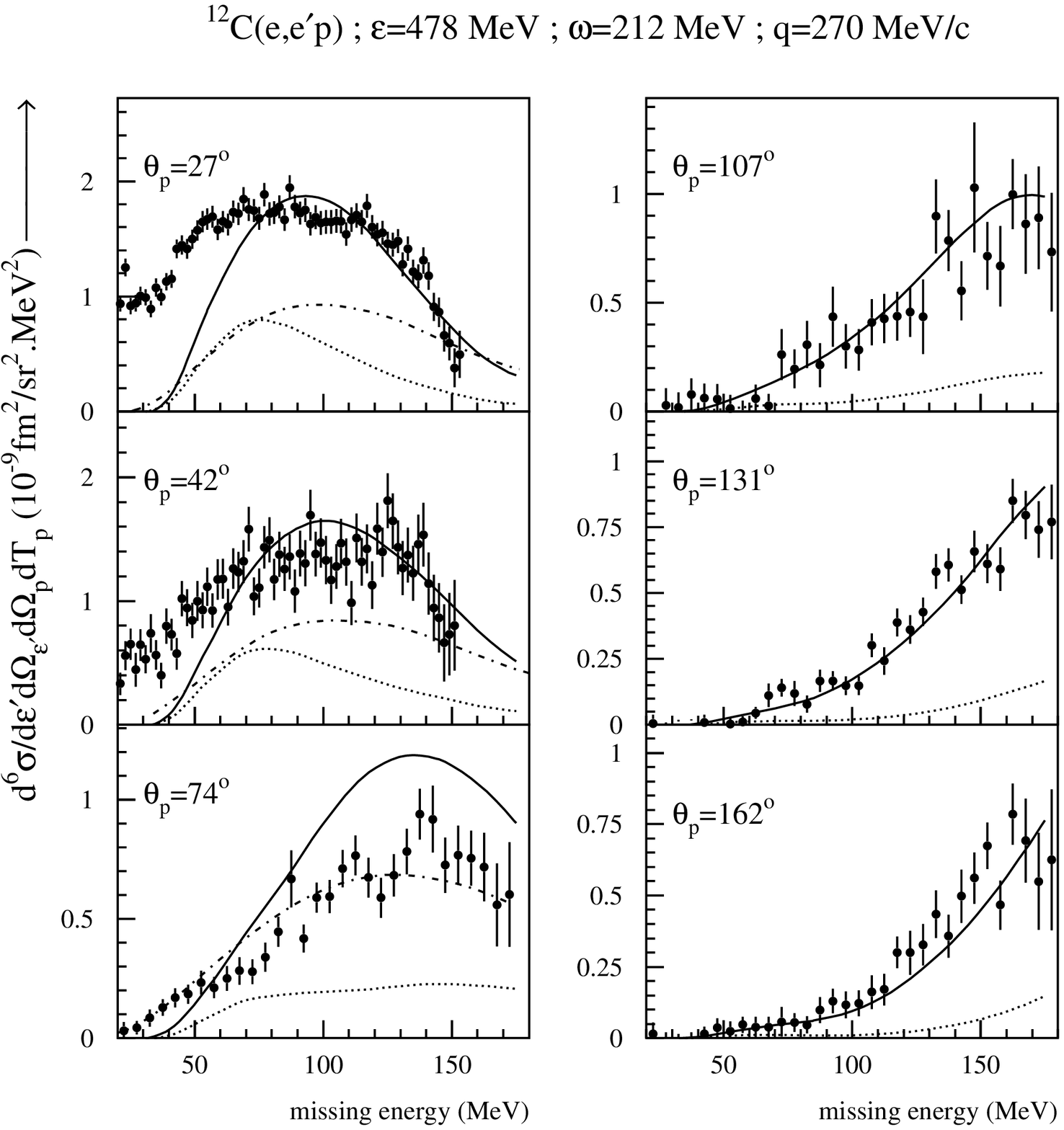}
\caption{Calculated contribution from the two-nucleon breakup channel to
the semi-exclusive $^{12}$C(e,e$'$p) reaction for $\epsilon$=478~MeV,
$\omega$=212~MeV and q=270~MeV/c.  For the dotted line only the SRC
current is retained. The dot-dashed line represents the result when only
accounting for the pion degrees of freedom (including intermediate
$\Delta _{33}$ creation).  The solid line is obtained
when coherently adding the short-range and mesonic contributions.  The data are
from Ref.~\protect\cite{leon}. The proton angles are expressed
relative to the direction of the momentum transfer $\protect\vec{q}$.
For the forward proton angles (27,42 and 74$^o$) the azimuthal angle
$\phi_p=180^o$, for the other angles $\phi_p=0^o$.}
\label{semiexclu}
\end{figure}

Although two-nucleon knockout reactions remain obvious choices to study
nucleon-nucleon correlations, in what follows we illustrate that
(e,e$'$p) studies can
be used to test theories of SRC when
performed under carefully selected kinematics.
In particular, we will concentrate on
the semi-exclusive (e,e$'$p) reaction performed at high missing momenta
($\vec{p}_m=\vec{p}_p-\vec{q}$) and energies ($E_m = \omega - T_p - T_{A-1}$).
Results of calculations for the semi-exclusive $^{12}$C(e,e$'$p) reaction
including short-range effects, pion-exchange
and isobaric currents are shown in Fig.~\ref{semiexclu}.  The
kinematics are from Ref.~\cite{leon} representing the first
semi-exclusive (e,e$'$p) measurement
on a complex nucleus for a whole range of proton angles.
The calculations have been performed along the lines explained in
Refs.~\cite{ryc94,ryc95}.  The model assumes that the
semi-exclusive strength above the two-nucleon emission threshold
arises from two-nucleon knockout.  Consequently, the calculations
involve an
integration over phase space of the undetected particle (either a proton
or a neutron) and a sum over all possible quantum
states $\mid J_R M_R>$ of the
residual nucleus.  Accordingly, the semi-exclusive (e,e$'$p) cross
sections are determined by two-nucleon knockout
matrix elements of the type (\ref{eq:trans}).
Just as for the (e,e$'$NN) channel the relative importance of the
different types of correlations can be investigated by retaining a
selected number of terms in the interaction hamiltoninan of
Eq.~(\ref{effh}).
Looking at Fig.~\ref{semiexclu} it becomes clear that the SRC
contribution (dotted line) represents a sizeable contribution
of the measured strength at the forward proton angles.  At backward proton
angles, the SRC contribution becomes marginal and the calculated
cross section is dominated by the pionic and isobaric degrees of
freedom.  As such, the data at large $\theta _p$'s are ideal to gauge the
meson-exchange part of the cross section.  As the direction of proton
detection moves closer to $\vec{q}$, it is not but after coherently
adding the short- and long-range terms in the effective interaction
hamiltonian that a reasonable overall description of the data can be
obtained.

It is worth investigating the driving mechanism behind the
proton angle dependence in the relative importance of the different
photoabsorption mechanisms.
As will become clear in the course of this paragraph,
a natural explanation of the qualitative features of the SRC cross
sections in Fig.~\ref{semiexclu} is provided by the two-nucleon
correlation model (TNC) as developed by Ciofi degli Atti {\em et al.}
\cite{ciofi1,ciofi2}.  In the most naive version of the TNC model it is
assumed that the short-range correlations in finite nuclei are governed
by those configurations in which the high momentum of a bound nucleon
${\vec p}_m$ is balanced by a a momentum  $-{\vec p}_m$ of a correlated
nucleon and that the remaining (A-2) nucleons act as spectators in this
process.  Discarding all effects related to the FSI this picture
predicts the following relation between missing energy and momentum in
case of emission of a correlated nucleon \cite{ciofi2,newref} :
\begin{equation}
E_m \approx S_{NN}+<E_x^{hh'}> + \frac{(A-2)p_m^2}{2(A-1)M_N} \;,
\label{eq:tnc}
\end{equation}
where $S_{NN}$ is the separation energy for two-nucleon knockout and
$<E_x^{hh'}>$ the average excitation energy of the two-hole state
$\mid h^{-1} h'^{-1}>$ in which the A-2 system is created (in the
$^{12}$C case reasonable values are
$<E_x^{(1p)^{2}}>$=0~MeV, $<E_x^{(1p1s)}>$=25~MeV and
$<E_x^{(1s)^{2}}>$=50~MeV).   It is worth mentioning that the above
relation (\ref{eq:tnc}) is based on purely kinematical arguments. When
applied to the kinematics of Fig.~\ref{semiexclu} the above relation
predicts a maximized likelihood to eject correlated protons for
$E_m \approx 85~MeV (\theta_p=27^o)$,
$E_m \approx 95~MeV (\theta_p=42^o)$,
$E_m \approx 130~MeV (\theta_p=74^o)$,
$E_m \approx 155~MeV (\theta_p=107^o)$,
$E_m \approx 165~MeV (\theta_p=131^o)$ and
$E_m \approx 170~MeV (\theta_p=162^o)$.   This observation explains why
the peaks of the dotted curves shift to higher $E_m$ with increasing
proton angle.
Another striking feature of the dotted curves in Fig.~\ref{semiexclu}
is that SRC effects loose in importance as the proton angle becomes larger.
This can be explained by considering that for the backward proton angles
very large missing momenta are probed and that the probability to find a
nucleon is a steadily decreasing function with growing momentum.  From
the above discussion it is clear that our dynamic and microscopic model
to deal with the SRC effects seems to reproduce the trends predicted by
the TNC model that is based on purely kinematical grounds.

Summarizing, we have developed a microscopic model that aims at
estimating the quantitative effect of SRC on (e,e$'$pp) and (e,e$'$pn)
cross sections in complex nuclei.  In our approach, both the effects of
the SRC in the nuclear wave functions and the pion-exchange currents in
the photoabsorption mechanism are  treated. The latter are shown to be
highly competitive with the SRC and necessary to obtain a reasonable
agreement with the semi-exclusive $^{12}$C(e,e$'$p) data of a recent
NIKHEF-K experiment.  Even for the (e,e$'$pp) cross sections the effect
of the SRC was found to be of similar magnitude than the strength
created by intermediate $\Delta _{33}$ creation.  Further, our
microscopic approach confirms that semi-exclusive (e,e$'$p) reactions
can be very useful tools to test different theories of SRC, provided
that they are performed under suitable kinematics.

This work was supported by the National Fund for Scientific
Research (NFWO).


\begin{thebibliography}{99}
\bibitem{ryc94}
J. Ryckebusch, L. Machenil, M. Vanderhaeghen, V. Van der Sluys and M.
Waroquier, Phys. Rev. {\bf C49} (1994) 2704.
\bibitem{benhar0}
O. Benhar, A. Fabrocini, S. Fantoni and I. Sick, Nucl. Phys. {\bf A579}
(1994) 493.
\bibitem{benhar}
O.Benhar, A. Fabrocini and S. Fantoni, in  {\em Modern Topics in Electron
Scattering}, eds. B. Frois and I. Sick (World Scientific, Singapore, 1991),
460.
\bibitem{revmod}
O. Benhar, V.R. Pandharipande and S.C. Pieper, Rev. Mod. Phys. {\bf 65}
(1993) 817.
\bibitem{pieper}
S.C. Pieper, R.B. Wiringa and V.R. Pandharipande, Phys. Rev. {\bf C46}
(1992) 1741.
\bibitem{weise}
W. Weise, M.G. Huber and M. Danos, Z. Phys. {\bf 236} (1970) 176.
\bibitem{leideman}
W. Leidemann, G. Orlandini, M. Traini and E.L. Tomusiak, Phys. Rev. {\bf
C50} (1994) 630.
\bibitem{marc}
M. Vanderhaeghen, L. Machenil, J. Ryckebusch and M. Waroquier, Nucl.
Phys. {\bf A580} (1994) 551.
\bibitem{co}
G. Co$'$, A. Fabrocini, S. Fantoni and I.E. Lagaris, Nucl. Phys. {\bf
A549} (1992) 439.
\bibitem{ryc93}
J. Ryckebusch, M. Vanderhaeghen, L. Machenil and M. Waroquier, Nucl.
Phys. {\bf A568} (1994) 828.
\bibitem{carfsi}
C. Giusti, F.D. Pacati and M. Radici, Nucl. Phys. {\bf A546} (1992) 607.
\bibitem{giusti1}
C. Giusti and F.D. Pacati, Nucl. Phys. {\bf A535} (1991) 573.
\bibitem{clark}
J.W.~Clark, Lecture Notes in Physics {\bf 138}, eds. R. Guardiola and J.
Ros (Springer-Verlag, Berlin, 1981), 184.
\bibitem{leon1}
L.J.H.M. Kester {\em et al.}, Nucl. Phys. {\bf A553} (1993) 709c.
\bibitem{giusti2}
C. Giusti and F.D. Pacati, Nucl. Phys. {\bf A571} (1994) 694.
\bibitem{leon}
L.J.H.M. Kester {\em et al.}, Phys. Lett. {\bf B344} (1995) 79 {\em and}
Ph.D. thesis, Free University of Amsterdam, 1993, unpublished.
\bibitem{ryc95}
J. Ryckebusch, V. Van der Sluys, M. Waroquier, L.J.H.M. Kester, W.H.A.
Hesselink, E. Jans and A. Zondervan, Phys. Lett. {\bf B333} (1994) 310.
\bibitem{ciofi1}
C. Ciofi degli Atti, S. Simula, L.L. Frankfurt and M.I. Strikman, Phys.
Rev. {\bf C44} (1991) R7.
\bibitem{ciofi2}
C. Ciofi degli Atti, S. Liuti and S. Simula, Phys.
Rev. {\bf C41} (1991) R2474.
\bibitem{newref}
J.M. Le Goff {\em et al.}, Phys. Rev. {\bf C50} (1994) 2278.
\end{thebibliography}
\end{document}